# Influence of heat treatment excursion on critical current and residual resistivity ratio of ITER Nb₃Sn strands

J. Lu*, D. R. McGuire, S. Hill, and R. Niu, National high magnetic field laboratory, Tallahassee, Florida 32310 USA

K. Chan and N. N. Martovetsky, Oak Ridge National Laboratory, Oak Ridge, Tennessee 37831 USA

*Email: junlu@magnet.fsu.edu

**Abstract**

Heat treatment is critically important to the performance of Nb$_3$Sn superconducting strands. For very large Nb$_3$Sn magnet coils, such as the International Thermonuclear Experimental Reactor (ITER) Central Solenoid (CS) coils, heat treatment carries risk of temperature and time excursion, which may result in performance degradation. Therefore, it is prudent to study the effect of possible excursion on Nb$_3$Sn performance. In this study, Nb$_3$Sn strands used for ITER CS coils are heat treated at different temperatures for different times. Their critical current, residual resistance ratio and hysteresis losses are measured. It is found that in the range we studied, critical current and hysteresis losses do not change significantly. Residual resistance ratio, however, decreases with increasing heat treatment temperature and time. This is attributed to the diffusion of metallic elements from the plated Cr layer to the copper stabilizer. Based on a model of metallic elements diffusion, a numerical code is developed to predict residual resistance ratio as a function of heat treatment temperature and time.

**Keywords:**   Nb$_3$Sn, Heat treatment, critical current, RRR, Cr diffusion

1. Introduction

Nb$_3$Sn strand is a composite of Cu, pure Nb, and Sn or Sn containing components, which requires a reaction heat treatment to form A15 phase of superconducting Nb$_3$Sn. Obviously the superconducting properties of a Nb$_3$Sn strand is strongly influenced by heat treatment schedule. In a heat treatment, Nb$_3$Sn reaction is limited by diffusion of Sn to Nb filaments [1]. So as the heat treatment advances, thickness of the reacted A15 phase increases with temperature and time. At the same time, Sn content of the A15 layer increases. Both lead to increase in critical current ($I_c$). Heat treatment at too high of a temperature for too long, however, results in large $Nb_3Sn$ grain size with fewer grain boundaries. This reduces the flux pinning ability by grain boundaries and consequently lowers critical current. In order to maximize A15 layer thickness and its Sn content while minimizing its grain size, heat treatment schedule is optimized for specific strand manufacturing route and its architecture. Many studies have been carried out [2]-[16] on the optimization of $Nb_3Sn$ heat treatment schedule to bring out the best possible superconducting properties. In principle, deviation from the optimized heat treatment schedule may lead to performance degradation. For large $Nb_3Sn$ magnet coils, due to the challenges in achieving uniform temperature of large thermal mass, heat treatment temperature and time excursions are always possibilities, which might lead to coil performance degradation.

The Central Solenoids (CS) of the International Thermonuclear Experimental Reactor (ITER) are wound by $Nb_3Sn$ cable-in-conduit conductors and need to be heat treated after coil-winding [17]. The coil of such large size requires a very large heat treatment facility [18] which makes precise and uniform temperature control very challenging. In addition to the possible heat treatment interruption due to electrical power or other failures, temperature and time excursion during heat treatment are also possible. In order to minimize the risk of coil performance degradation, therefore, it is prudent to study the effect of heat treatment temperature and time excursion from the optimized heat treatment schedule.

Moreover, it has been reported that Residual Resistance Ratio (RRR) of Cr plated ITER strands decrease with increasing heat treatment time [19-23]. This has been attributed to Cr diffusion into copper stabilizer. Nevertheless, despite of continuing efforts of past decades by a number of research groups, a complete understanding of this phenomenon and ability of predicting strand RRR by its heat treatment schedule are still lacking. In this paper, we studied the effect of heat treatment temperature and time excursion on $Nb_3Sn$ critical current ($I_c$) and RRR. Our discussion is focused on the prediction of RRR of Cr plated strands based on its heat treatment schedule.

2. Experimental

The $Nb_3Sn$ strands used in this study are made by Japan Superconductor Technology Inc. (JASTEC) and Kiswire Advanced Technology (KAT). JASTEC and KAT strands are made by bronze and internal-tin route respectively. Both are Cr plated with nominal diameters of 0.82 mm and copper/non-Cu ratio of 1. Figure 1 shows their cross-sections.

For this study, heat treatments were divided into two steps as listed in Table I. For each strand, 36 sets of samples (each set includes one $I_c$, one RRR, and one hysteresis losses sample) were heat treated together in step-one where temperature ramp rate to all levels was 5 C/h. Then samples were furnace cooled to room temperature. For each of the step-two heat treatment, temperature was ramped up at 60 C/h to a level and held for a duration. Combinations of three temperatures and three or more durations were designed for step-two heat treatments. Up to four samples were heat treated for each combination. After step-two heat treatment, temperature was ramped down to 500 C at 5 C/h followed by furnace cool to room temperature. Total of 13 and 9 step-two heat treatments were performed for JASTEC and KAT strand respectively as shown in Table I. The manufacturer optimized step-two heat treatment schedule are 100 h at 650 C for JASTEC strand and 120 h at 650 C for KAT strand. All heat treatments were carried out in flowing argon.

After heat treatment, samples were tested for $I_c$, hysteresis losses ($Q_{hyst}$) and RRR. $I_c$ were measured at 4.2 K in 11, 12, and 13 T magnetic fields; a criterion of 10 µV/m was used to determine $I_c$. RRR is defined as the resistance ratio between 273 K and 20 K. $Q_{hyst}$ during ± 3 T field sweep were measured by a vibrating sample magnetometer at 4.22 K. For most heat treatment temperature and time combinations, multiple samples were heat treated and tested. In these cases, average property values are presented in this paper. More details of measurement techniques can be found in [24]. Cross-section of some heat treated samples were polished and examined by Zeiss 1540 Field Emission Scanning Electron Microscope (FESEM). The backscattered electron intensity in the FESEM is sensitive to atomic number of an element, so Sn exhibits brighter contrast than Cu or Nb. This is conveniently used to detect Sn rich region in $Nb_3Sn$ filaments.

### 3. Results and discussions

#### a. Cross-sections of reacted strands by FESEM

Figures 2(a) – 2(d) show cross-sections of JASTEC samples heat treated for 0, 10, 40, and 100 h at 650 C. These filaments are about 2 μm in diameter with some minor bridging between filaments. As heat treatment progressed from 0 to 100 h, the Sn content of each filament gradually increases starting from the periphery of the filament. For the sample reacted for 40 h (figure 2(c)), a Sn deficient core is still clearly visible at the center of the filament. Whereas at 100 h (figure 2(d)), the reaction seems completed as indicated by the absence of the Sn deficient core.

#### b. Critical current and hysteresis losses

Figures 3(a) and 3(b) show 12 T $I_c$ versus heat treatment temperatures for JASTEC and KAT samples respectively. Evidently $I_c$ is not sensitive to heat treatment temperature or time within the experimental uncertainty which can be up to 5% of $I_c$ [25]. When the heat treatment time in step-two is reduced to 20 h or less, however, $I_c$ of the JASTEC strand reduces significantly as shown in figure 4 (a) where data can be fitted (solid line) using an exponential function

$$I_c = I_{c0} + A[1 - exp(-t/\tau)] \quad (1)$$

Where $t$ is heat treatment time, $I_{c0}$, $A$, and a time constant $\tau$ are fitting parameters. The best fit of $\tau = 12$ h suggests a relatively rapid reaction process. This exponential approach to an asymptotic level, as described by equation (1), is consistent with a diffusion controlled $Nb_3Sn$ growth where the growth rate decreases as Sn in the bronze matrix is being depleted.

Figure 4(a) also shows that even before the step-two heat treatment (at time = 0 h in the figure), $I_c$ is already appreciable at about 75% of the saturated value, thanks to 250 h at 570 C in step-one. After that, for only 20 h at 650 C, the $I_c$ reaches over 96% of the saturated value. Therefore the manufacturer optimized schedule of 100 h at 650 C for JASTEC strand is sufficiently long to avoid $I_c$ degradation. Upper critical field $B_{c2}^*$(4.2 K) versus heat treatment time at 650 C is plotted in figure 4(b). Here $B_{c2}^*$(4.2 K) values are obtained from $I_c(B)$ by linear extrapolation of Kramer plots [26] (self-field correction is not applied to $I_c(B)$ data). An exponential function similar to equation (1) also provides good fits for $B_{c2}^*$(4.2 K) versus time with a similar time constant of $\tau = 13$ h. Since $B_{c2}^*$(4.2 K) is related to Sn content in A15 phase [27], the rapid $B_{c2}^*$(4.2 K) rise in the beginning of step-two heat treatment indicates a quick increase in Sn content of A15, in addition to the increase in A15 layer thickness. It should be noted that, although minor under reaction does not cause significant $I_c$ degradation, it might cause considerable $I_c$ strain sensitivity which is detrimental to the performance of $Nb_3Sn$ under high stress [3].

Since hysteresis losses of a superconducting strand ($Q_{hyst}$) is proportional to its magnetization, and its magnetization is proportional to its $I_c$, $Q_{hyst}$ is directly linked to $I_c$. Therefore, it is expected that $Q_{hyst}$ increases with heat treatment time in the way analogous to $I_c$ as shown in figure 5. The solid line in the figure is a fit by a function similar to equation (1) with a time constant of $\tau = 12$ h.

   c. *RRR experimental results and calculation*

It is known that RRR of Cr plated strands decreases with increasing heat treatment time [19-23]. In order to verify that Cr layer is the main contributor to RRR degradation, the Cr layer is removed from a set of JASTEC samples by chemical etching with 37% HCl. These samples were heat treated together with a few Cr plated samples at 570 C for various durations. As shown in figure 6, RRR of samples without Cr have rather weak heat treatment time dependence; whereas RRR of samples with Cr shows a clear downward trend with increasing heat treatment time. This experiment demonstrates very clearly that plated Cr layer is largely responsible for the observed RRR degradation.

As for the effect of step-two heat treatment, RRR of JASTEC samples is plotted against heat treatment temperature and time and in figure 7 which show a clear trend of degradation with increasing heat treatment temperature and time. This degradation in RRR is consistent with previously observed phenomenon which was attributed to the Cu contamination due to the diffusion of metallic elements in plated Cr. As depicted schematically in figure 8, Cr concentration distribution in Cu stabilizer at distance $x$ from Cr/Cu interface and at time $t$ can be calculated by a complementary error function [19],

$$C(x,t) = C_0 \left[ \text{erfc}\left(\frac{x}{2(Dt)^{1/2}}\right) \right] \qquad (1)$$

Where $C_0$ is the initial Cr concentration at the Cr/Cu interface which, in this case, equals the solubility of Cr in Cu at the heat treatment temperature, $D$ is the diffusion coefficient which follows an Arrhenius equation,

$$D = D_0 \exp(-E_a/kT) \qquad (2)$$

Where $D_0$ is a constant, $E_a$ is an activation energy, and $k$ is Boltzmann constant. For calculation of resistivity of contaminated copper, the resistivity $\rho$ can be related to low level impurities concentration $C(x, t)$ by Matthiessen's rule,

$$\rho(x, t) = \rho_0 + \alpha C(x, t) \qquad (3)$$

Where $\rho_0$ is the resistivity of pure copper at a certain temperature, $\alpha$ is a constant which represents resistivity-increase per atomic percent of impurity. In RRR calculation, for convenience we consider Nb$_3$Sn strand as Cu stabilizer region and non-Cu core region connected in a parallel circuit. So RRR can be calculated by,

$$RRR = \frac{K_{Cu-20K} + K_{Core-20K}}{K_{Cu-273K} + K_{Core-273K}} \qquad (4)$$

Where $K_{Cu-20K}$ and $K_{Cu-273K}$ are conductances of Cu stabilizer at 20 and 273 K respectively. $K_{Core-20K}$ and $K_{Core-273K}$ are conductances of the non-Cu core at 20 and 273 K respectively, which can be measured experimentally.

Using equations (1) - (3), we can calculate the conductance of Cu stabilizer $K_{cu}$ by radially integrating conductance over the entire Cu region,

$$K_{Cu} = \int_{r_1}^{r_2} \frac{2\pi r dr}{\rho(r)} = \int_{r_1}^{r_2} 2\pi r [\rho_0 + \alpha C_0 erfc(\frac{r_2-r}{2\sqrt{Dt}})]^{-1} dr \qquad (5)$$

Where $D$ is a function of temperature follows equation (2), $r_1$ and $r_2$ are inner and outer radii of the Cu stabilizer respectively as depicted in figure 8.

We developed a Microsoft Excel VBA code based on (4) and (5) to numerically calculate RRR of a Cr plated strand as a function of heat treatment temperature and time. Parameters used in the code for JASTEC strand are listed in Table II. Some of these parameters are explained as follows. The value of $\alpha = 4.0 \times 10^{-8}$ Ω-m/at% is based on experimental result by Linde [28]. Similar value is published by Gregory *et al.* [29]. This value was also adopted by Novosilova *et al.* [22] in their RRR simulation. Solubility of Cr in Cu, $C_0$, has considerable uncertainty according to a critical review by Chakrabarti and Lauphlin [30]. Due to this considerable uncertainty, we ignore its temperature dependence for simplicity, use 0.1 at% for 640 - 660 C temperature range based on data in [30]. It is also consistent with $C_0$ value used by Novosilova *et al.* [22] in their RRR calculations. Initial Cu RRR, $D_0$ and $E_0$ as free parameters which are adjusted to best fit experimental data in figure 7. The calculated RRR versus heat treatment time for all three temperatures are plotted in figure 7 as solid lines. The fitted values of these free parameters are listed in Table II as well.

In the following, we discuss the significance of the values of free parameters Initial Cu RRR, $D_0$ and $E_0$. For simplicity, the value of initial Cu RRR takes into account the effect of step-one heat treatment which is therefore not calculated separately. When the fitted initial Cu RRR of 540 is used to calculate the RRR of the strand, which includes both Cu and non-Cu regions, a strand RRR of 347 is obtained. This is in a reasonable agreement with the RRR measured after step-one heat treatment as shown in figure 6. $D_0$ and $E_a$ obtained from fitting experimental data are compared with those in [31] and [32] and listed in table III. It is evident that the diffusion coefficients obtained in this work is significantly higher than those in [31] and [32]. If we considered the possible precipitation of Cr during cooling from heat treatment temperature, as suggested by Alknes *et al.* [23], to cause the measured low RRR, the diffusion coefficient would have been even higher. This discrepancy was also observed by Novosilova *et al.* [22] who suggested that diffusion of other elements that co-exist in the Cr coating, such as oxygen, might be the cause. However, as Alknes *et al.* [23] pointed out, oxygen as a main cause of RRR degradation is not likely. This is because the diffusion coefficient of O in Cu (also listed in Table III) is about 6 orders of magnitude greater than that of Cr in Cu. So RRR degradation would have been saturated very shortly after the start of heat treatment. In fact impurities of other metallic elements such as Ni, Fe, and Zn are commonly found in Cr plating electrolyte [34]. This presents another possible cause of RRR degradation. For example, Ni is completely soluble in Cu at $Nb_3Sn$ heat treatment temperatures, and its diffusion coefficient in Cu [35] is higher than that of Cr in Cu as shown in Table III. Therefore even a low level of 0.1 at% Ni in Cr layer would be sufficient to cause significant contamination of Cu, despite the fact that the effect of Ni impurity on Cu resistivity is somewhat smaller than that of Cr [28], [29]. Therefore, we speculate that the combined diffusions of Cr and other metallic impurities in the Cr layer result in RRR degradation which advances with heat treatment time and temperature. Unfortunately, this low level of metallic impurities in Cr layer is difficult to measure by conventional characterization techniques such as energy dispersive x-ray spectroscopy (EDS). Ultimately the complete understanding of this RRR degradation should be based on accurate measurement of Cr and other impurity content in Cu stabilizer as a function of distance from the Cr/Cu interface. This is a very challenging measurement, due to extremely high detection sensitivity needed. Previous measurements by laser mass spectrometer [20], and high sensitivity EDS [23] only had limited success.

In order to demonstrate the validity of our RRR calculation code, a few additional sets of RRR versus heat treatment time data are plotted in figure 9. These include RRR of KAT strands at 640, 650, and 660 C, as well as some data reproduced from the literature [21]-[24] where heat treatments were at 650 C only. A downward trend of RRR with heat treatment time is evident. The same calculated curves as presented in figure 7 is also plotted in figure 9. Despite the differences in strand manufacturer, testing labs and the step-one heat treatment schedule, a universal trend of RRR versus heat treatment time is clear. It suggests the predictability of this phenomenon, and our RRR calculation code with the parameters listed in Table II seems to be very useful in predicting RRR of Cr plated strands.

4. Conclusion

Heat treatment sensitivity of JASTEC and KAT strands for ITER CS coils are studied by measuring $I_c$, $Q_{hyst}$ and RRR of samples heat treated at various temperatures and times. We found that moderate variation in temperature and time of the step-two heat treatment does not significantly change $I_c$. In the case of JASTEC strand, after the step-one heat treatment, samples already have about 75% of its maximum $I_c$. RRR decreases with increasing heat treatment temperature and time due to the diffusion of contaminants from Cr layer into the Cu stabilizer. The RRR of a Cr plated $Nb_3Sn$ strand can be predicted by a Cr diffusion model as demonstrated by our calculation in this paper.

5. Acknowledgement

The authors from NHMFL thank financial supports in part by the U.S. Department of Energy vie US-ITER under subcontract 4000110684, by the National Science Foundation under Grant DMR-0084173, and by the State of Florida.

The authors from US-ITER would like to thank the U.S. Department of Energy by Lawrence Livermore National Laboratory under Contract DE-AC52-07NA27344, by UT-Battelle, LLC, under contract DE-AC05-00OR22725 with the U.S. Department of Energy, and by the U.S. Department of Energy, Office of Science, Office of Fusion Energy Sciences.

## 7. Figure caption

Table I Heat treatment schedules

Table II Parameters used in the RRR calculation.

Table III Diffusion coefficients

Figure 1 Cross-section of $Nb_3Sn$ wires measured in this paper, left JASTEC, right KAT.

Figure 2 Scanning electron microscopy images of cross-section of JASTEC strand heat treated at 650 C for (a) 0 h, (b) 10 h, (c) 40 h, and (d) 100 h.

Figure 3 Critical current versus heat treatment temperature for a) JASTEC strands and b) KAT strands.

Figure 4 JASTEC strand data. (a) $I_c$ as function of heat treatment time at 650 C in 11, 12, and 13 T fields. The solid lines are simulations as described by equation (1) with $\tau = 12$ h. (b) Upper critical field $B_{c2}^*$(4.2 K) obtained by Kramer method as a function of heat treatment time at 650 C. The solid line is a fit by a function similar to equation (1) with $\tau = 16$ h.

Figure 5 hysteresis losses as a function of heat treatment time. The solid line is a fit with a function similar to equation (1) where time constant $\tau = 12$ h.

Figure 6 RRR as function of heat treatment time at 570 C (first step) for samples with and without plated Cr.

Figure 7 RRR as a function of step-two heat treatment time at different temperatures. The solid lines are results of calculation for 640, 650, 660 C using parameters listed in Table II.

Figure 8 Compiled data of RRR versus heat treatment time. Solid lines are from numerical calculation, the same as in figure 7(a).

Table I. Heat treatment schedules

|          | JASTEC | KAT |
|----------|--------|-----|
| Step one | 570 C for 250 h | 210 C for 50 h, followed by 340 C for 25 h, followed by 450 C for 25 h, followed by 570 C for 100 h |
| Step two | 640 C for 80, 100, 150 h<br>650 C for 0, 10, 20, 40, 80, 100, 150 h<br>660 C for 80, 100, 150 h | 640 C for 100, 120, 150 h<br>650 C for 100, 120, 150 h<br>660 C for 100, 120, 150 h |

*After step one, samples were furnace cooled. Step two ramp up rate is 60 C/h. After step two heat treatment, temperature is ramped down to 500 C at 5 C/h before furnace cooled to room temperature.

**The ITER heat treatment schedule for JASTEC and KAT wire at step two is 650 C for 100 and 120 h respectively.

Table II. Parameters used in the RRR calculation.

| Parameters used in RRR simulation | | Remark |
|---|---|---|
| Diameter (mm) | 0.82 | Measured |
| Core diameter (mm) | 0.591 | Measured |
| $\rho_{Cu-273K}$ ($\Omega$-m) | $1.54 \times 10^{-8}$ | [6] |
| Initial Cu RRR | 540 | Free |
| $\rho_{Core-273K}$ ($\Omega$-m) | $2.73 \times 10^{-8}$ | Measured |
| Core RRR | 3.98 | Measured |
| Solubility limit (at%) | 0.1 | [30] |
| Resistivity-increase per 1 at% $\alpha$ ($\Omega$-m/at%) | $4.0 \times 10^{-8}$ | [28], [29] |
| Cr diffusion $D_0$ (m$^2$/s) | $4.2 \times 10^{-2}$ | Free |
| Cr diffusion activation $E_a$ (eV) | 2.53 | Free |

Table III Diffusion coefficients

| Source | $E_a$ (eV) | $D_0$ (m$^2$/s) | Diffusion coefficient ($\times 10^{-16}$ m$^2$/s) | | | |
|---|---|---|---|---|---|---|
| | | | 570 C | 640 C | 650 C | 660 C |
| Cr (this work) | 2.53 | $4.2 \times 10^{-2}$ | 0.31 | 4.54 | 6.43 | 9.05 |
| Cr [31] | 2.02 | $3.37 \times 10^{-5}$ | 0.275 | 2.32 | 3.07 | 4.03 |
| Cr [32] | 2.10 | $1.11 \times 10^{-4}$ | 0.310 | 2.84 | 3.79 | 5.04 |
| Ni [35] | 0.99 | $2.69 \times 10^{-10}$ | 3.21 | 9.14 | 10.5 | 12.0 |
| O [33] | 0.70 | $1.16 \times 10^{-6}$ | $7.58 \times 10^5$ | $1.59 \times 10^6$ | $1.75 \times 10^6$ | $1.92 \times 10^6$ |

*Data from the references are extrapolated values.

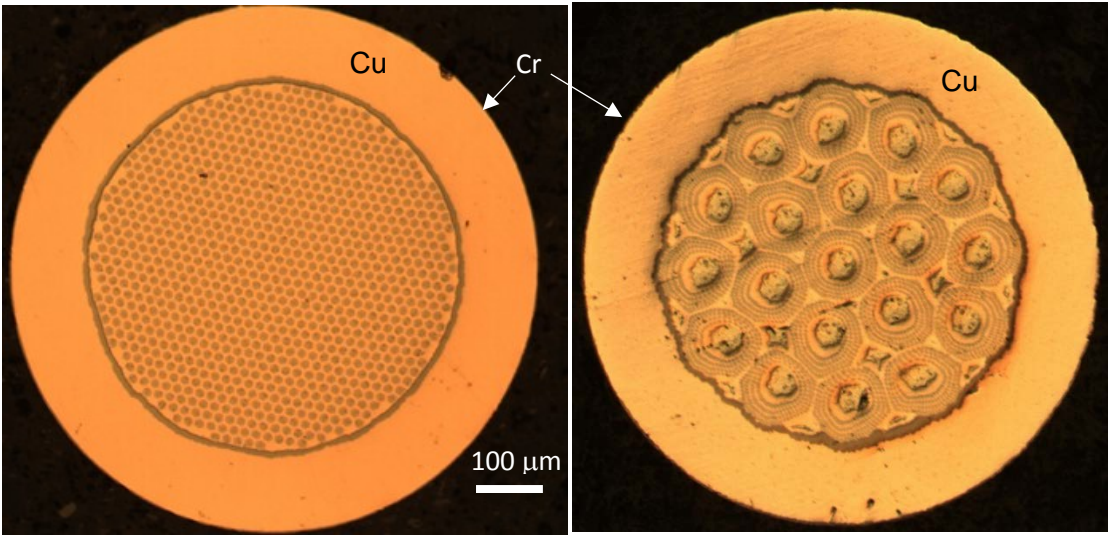

Fig. 1 J. Lu et al

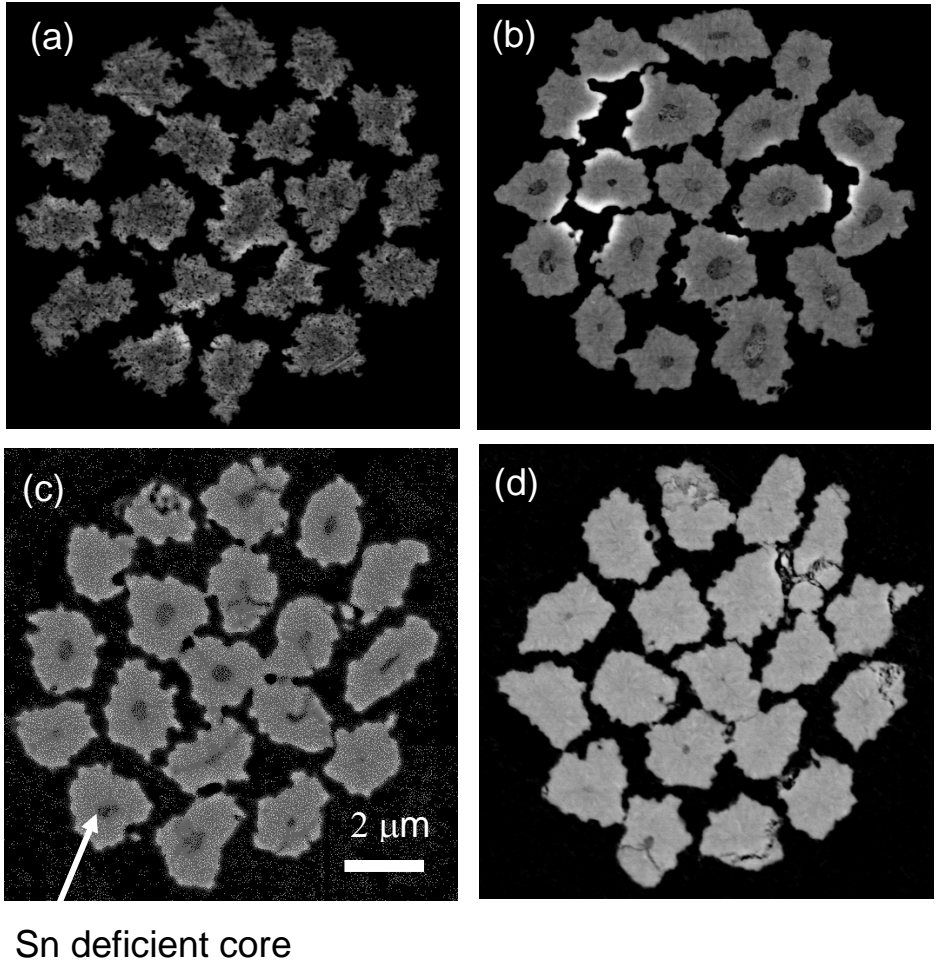

Sn deficient core

Fig. 2 J. Lu et al

(a)

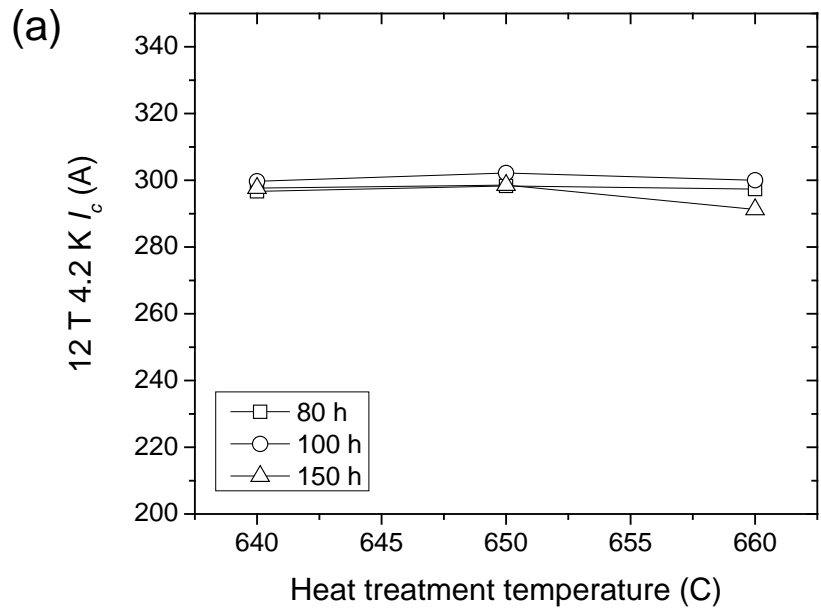

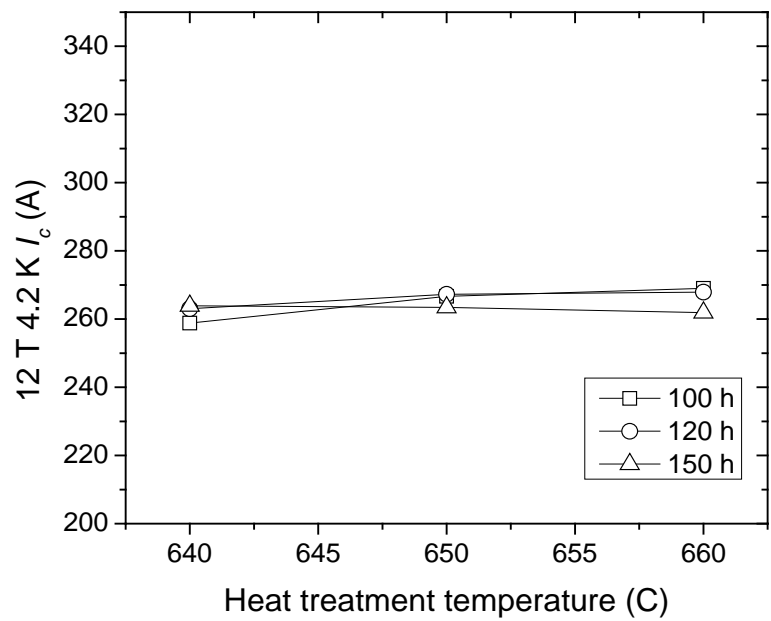

(a)

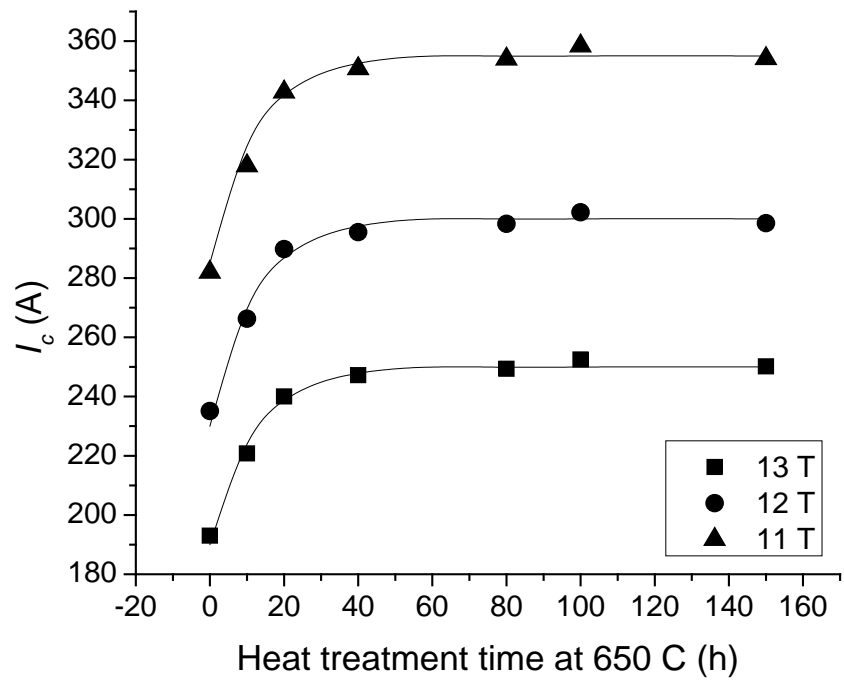

(b)

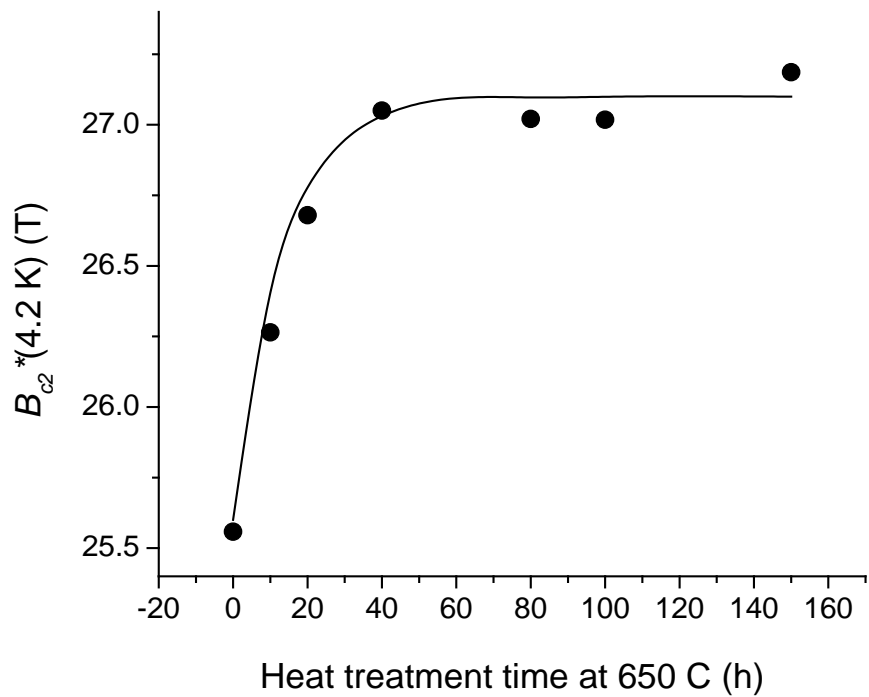

Fig. 4 J. Lu et al

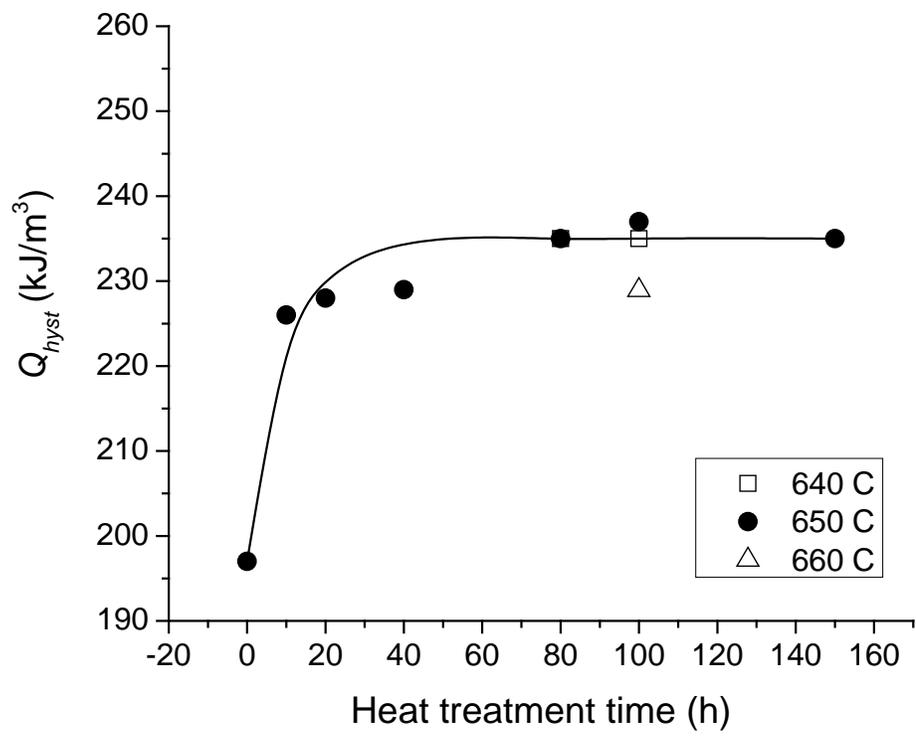

Fig. 5 J. Lu et al

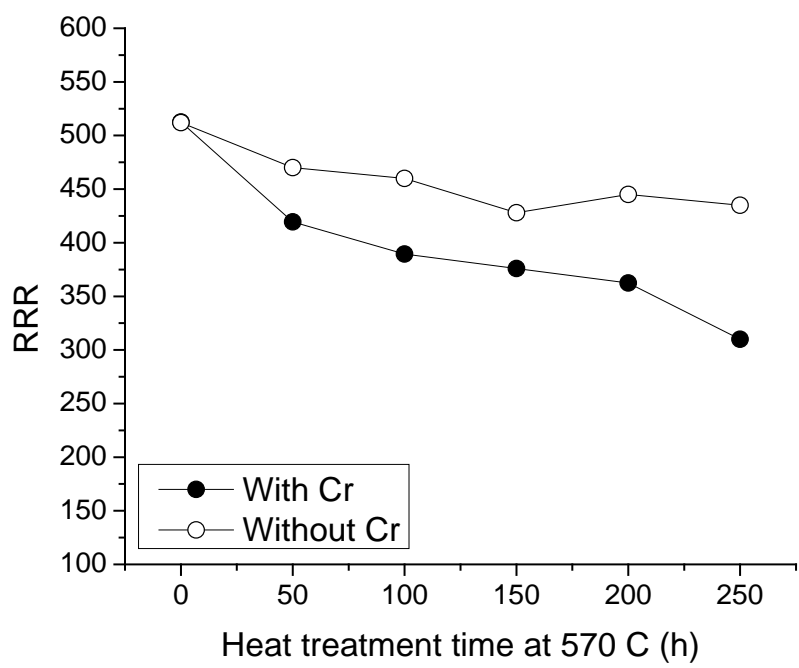

Fig. 6 J. Lu et al

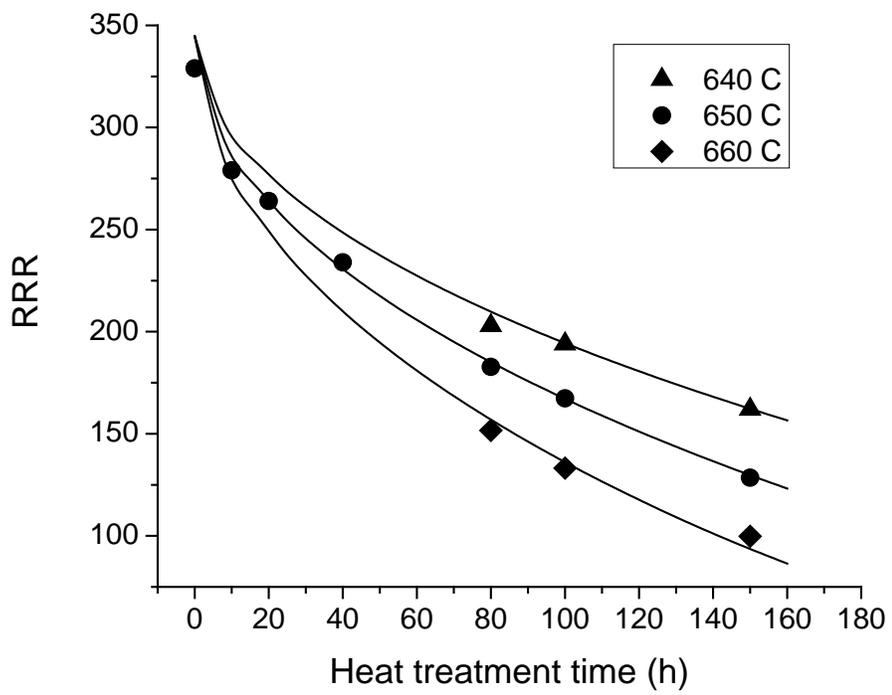

Fig. 7 J. Lu et al

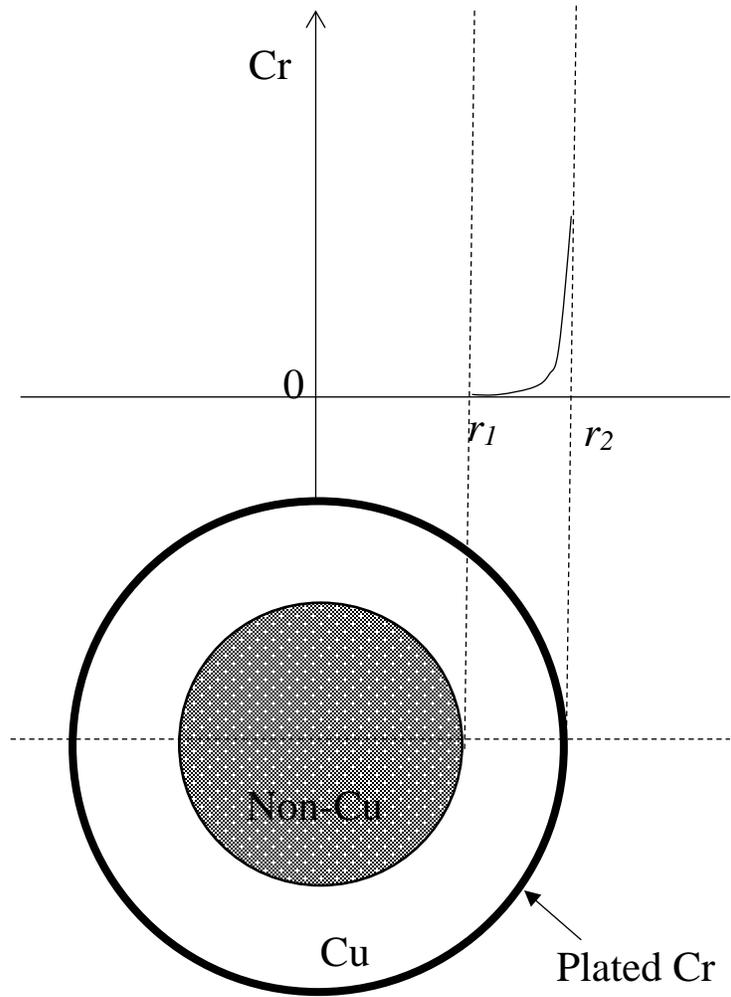

Fig. 8 J. Lu et al

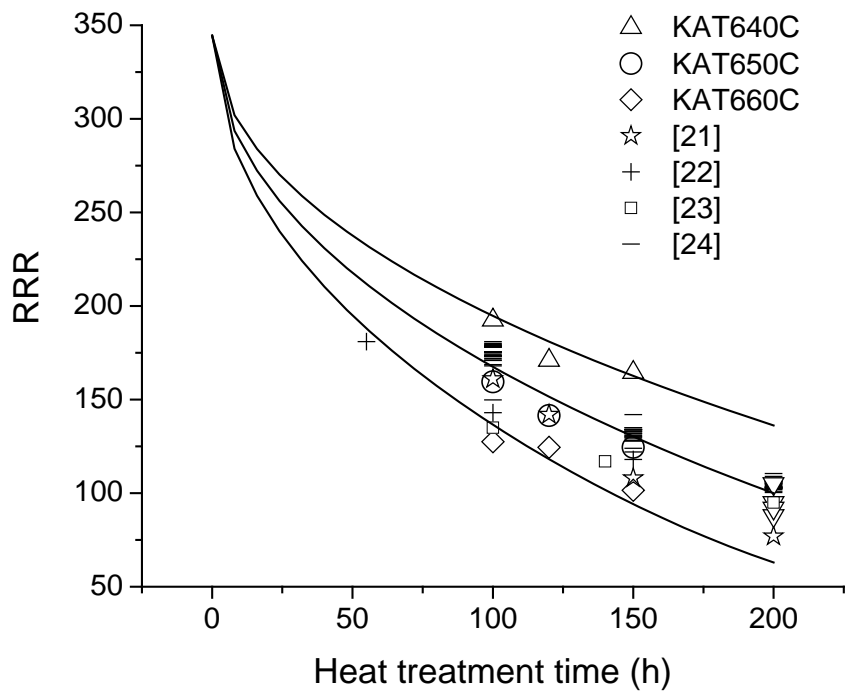

Fig. 9 J. Lu et al